\documentclass[useAMS,usenatbib,letterpaper]{mn2e}
\usepackage{graphicx}

\newcommand{\Ek}[1]{E_{{\rm k}, #1}}

\title[The Kelvin-Helmholtz instability in weakly ionised plasmas]{The
	Kelvin-Helmholtz instability in weakly ionised plasmas :
		Ambipolar dominated and Hall dominated flows.}
\author[A. C. Jones and T. P. Downes]{A.C. Jones$^{1,3}$ and 
	T. P. Downes$^{1,2,3}$\thanks{E-mail: turlough.downes@dcu.ie (TPD)}\\
$^{1}$School of Mathematical Sciences, Dublin City University,
	      Glasnevin, Dublin 9, Ireland\\
$^{2}$School of Cosmic Physics, Dublin Institute for Advanced
              Studies, 31 Fitzwilliam Place, Dublin 2, Ireland \\
$^{3}$National Centre for Plasma Science and Technology, Dublin
         City University, Glasnevin, Dublin 9, Ireland}
\begin{document}

\date{Accepted --. Received --; in original form --}

\pagerange{\pageref{firstpage}--\pageref{lastpage}} \pubyear{----}

\maketitle

\label{firstpage}

\begin{abstract}
The Kelvin-Helmholtz instability is well known to be capable of
converting well-ordered flows into more disordered, even turbulent,
flows.  As such it could represent a path by which the energy in, for
example, bowshocks from stellar jets could be converted into turbulent
energy thereby driving molecular cloud turbulence.  We present the results of 
a suite of fully multifluid magnetohydrodynamic simulations of this 
instability using the HYDRA code.  We investigate the behaviour of the
instability in a Hall dominated and an ambipolar diffusion dominated
plasma as might be expected in certain regions of accretion disks and 
molecular clouds respectively.

We find that, while the linear growth rates of the instability are unaffected 
by multifluid effects, the non-linear behaviour is remarkably different with 
ambipolar diffusion removing large quantities of magnetic energy while the Hall 
effect, if strong enough, introduces a dynamo effect which leads to continuing 
strong growth of the magnetic field well into the non-linear regime and
a lack of true saturation of the instability.
\end{abstract}


\section{Introduction}

The Kelvin-Helmholtz (KH) instability is an important instability
in almost any system involving fluids: it can occur anywhere that has a
velocity shear.  In astrophysical plasmas the KH instability can
provide the means of producing turbulence in a medium or the mixing of
material between two boundary layers.

The KH instability has been studied in a variety of astrophysical systems, 
from solar winds \citep{amerstorfer07, bettarini06, hasegawa04} and pulsar 
winds \citep{bucciantini06} to thermal flares \citep{venter06}. Due to its 
ability to drive mixing and turbulence, the KH instability has been considered 
relevant in protoplanetary disks \citep{johansen06, gomez05}, accretion disks 
and magnetospheres \citep{li04}, and other jets and outflows \citep{baty06}.  
Generally speaking, the assumptions of ideal magnetohydrodynamics (MHD) have 
been used in order to simplify the system of equations to be solved.
These assumptions are, however, not always valid.  Weakly ionised plasmas, for 
example, contain a large fraction of neutral particles as well as a number of 
charged particle fluids with differing physical characteristics.  Interactions 
between the various species can introduce non-ideal effects. Ambipolar 
dissipation and the Hall effect are two non-ideal effects that can greatly 
influence the development of the KH instability in a system by altering the 
dynamics of the plasma and the evolution of the magnetic field.  Astrophysical 
examples of such weakly ionised systems include dense molecular clouds 
\citep[e.g.][]{ciolek} and accretion disks around young stellar objects 
\citep[e.g.][]{wardle99disk}. In these systems, the relevant length scales are 
such that non-ideal effects can play an important role 
\citep{wardle04a, downes09, downes11}.

%The Kelvin-Helmholtz instability is of interest in molecular clouds.  The 
%exact source of the turbulent energy in these clouds remains unknown, although 
%one possible source is the interaction of protostellar jets with the 
%surrounding cloud, as first proposed by \citet{norman80}. The transfer of 
%momentum from a jet to the cloud is most likely to occur through
%so-called ``prompt entrainment'' \citep[e.g.][]{dyson84}. As the supersonic 
%protostellar jet propagates into the surrounding cloud, it forms a bow
%shock. This shock accelerates molecular cloud material, imparting
%momentum to it.  It is worth noting that the momentum imparted is relatively 
%well ordered: a further process must occur to convert this momentum into 
%turbulent motions.

Many authors have investigated the role of the KH instability in both
magnetised and unmagnetised astrophysical flows \citep[e.g.][]{frank96, mala96, 
hardee97num, downes98, keppens99}.  Most of these studies have investigated the 
KH instability in the context of either hydrodynamics or ideal MHD.  We
know that non-ideal effects are important in molecular clouds at length
scales below about 0.2\,pc \citep[e.g.][]{ois06, downes09} and hence it
is of interest to explore the KH instability in the context of either
non-ideal MHD or, preferably, fully multifluid MHD.  In more recent
years the emphasis of KH studies has been on including non-ideal
effects.  \citet{keppens99} studied both the linear growth and subsequent 
nonlinear saturation of the KH instability using resistive MHD numerical 
simulations. The inclusion of diffusion allowed for magnetic reconnection
and non-ideal effects were observed through tearing instabilities and  
the formation of magnetic islands.

\citet{palotti08} also carried out a series of simulations using
resistive MHD. They found that, following its initial growth, the KH
instability decays at a rate that decreases with decreasing plasma
resistivity, at least within the range of of resistivities accessible to
their simulations. They also found that magnetisation increases the
efficiency of momentum transport, and that the transport increases with
decreasing resistivity.  \citet{birk02} examined the case of a partially 
ionised dusty plasma, using a multifluid approach in which collisions could be 
included or ignored.  They found that collisions between the neutral fluid and 
dust particles could lead to the stabilisation of KH modes of particular 
wavelengths. The unstable modes led to a significant local amplification of 
the magnetic field strength through the formation of vortices and current 
sheets. In the nonlinear regime they observed the magnetic flux being 
redistributed by magnetic reconnection. It was suggested that this could be 
applicable to dense molecular clouds and have important 
implications for the magnetic flux loss problem \citep{umebayashi90}.

A comprehensive study was carried out by \citet{wiechen06} which
demonstrated the effect of dealing with the plasma using a multifluid 
scheme. This study focused on the effect of varying the properties of 
the dust grains. The results of the simulations led to the conclusions that 
more massive dust grains have a stabilising effect on the system while higher 
charged numbers have a destabilizing effect. It was found that there is no 
significant dependence on the charge polarity of the dust.

In a linear study of stellar outflows \citet{watson04} described how the 
charged and neutral fluids are affected differently by the presence of a 
magnetic field. This study is carried out using parameters chosen to reflect 
those of molecular clouds, and so is particularly relevant to our own study. 
The principal result of this paper is that for much of the relevant parameter 
space, neutrals and ions are sufficiently decoupled that the neutrals are 
unstable while the ions are held in place by the magnetic field. Since the 
magnetic field is frozen to the ionised plasma, it is not tangled by the 
turbulence in the boundary layer. The authors predict that with well-resolved 
observations, there should be a detectably narrower line profile in ionised 
species tracing the stellar outflow compared with neutral 
species, since ionised species are not participating in the turbulent 
interface with the ambient interstellar medium. The paper also includes a 
study of the growth rate of the instability.  It is found that at short 
length scales, the growth rate is well approximated by the growth rate of the 
hydrodynamic system. At larger scales and for super-Alfv\'{e}nic flows, 
the fastest growing mode is equal to that of the ideal MHD case.

\citet{shad08} carried out an analytical study of the Kelvin-Helmholtz 
instability in dusty and partially ionised outflows. They investigated 
primarily the effect of the presence of dust particles by varying their mass, 
charge and charge polarity. It was found that as the charge of the grain 
increased, the growth timescales also increased, implying a stabilising effect 
on the system. The stability of the system was also examined for dependence on 
the mass of the dust particles. It was found that for stronger magnetic 
fields, this did not affect the stability of the system. However, for weaker 
magnetic fields, the larger dust particles had a stabilising effect on the 
growing modes. This was in agreement with previous laboratory experiments 
\citep{luo01} and numerical simulations \citep{wiechen06}. Finally, as the 
magnetic field strength increased, the growth timescale of the unstable modes 
at a particular perturbation wavelength decreased.  By examining the 
combinations of the wavelength of the perturbation used, and the resultant 
growth timescales of the instability, \citet{shad08} concluded that the 
Kelvin-Helmholtz instability is a possible candidate for causing the formation 
of some of the physical structures observed in molecular outflows from young 
stars.

In this paper we perform numerical simulations of the complete evolution
of the Kelvin-Helmholtz instability in a weakly ionised, multifluid plasma 
including both its linear development, saturation and its subsequent
behaviour.  We include the physics of the Hall effect, ambipolar
diffusion and parallel resistivity in these simulations and we analyse
the role each of the former two effects on the development of the
instability.

The aim of this work is to investigate the influence of multifluid
effects on the growth and saturation of the KH instability.  In order to
develop a full understanding of the roles of the various non-ideal
effects, in particular the Hall effect and ambipolar diffusion, we run
simulations with parameters chosen to simulate very high, medium and
very low magnetic Reynolds number systems and with parameters chosen to ensure 
ambipolar-dominated flows and Hall-dominated flows.  We focus on
gaining an understanding of the general characteristics of the KH
instability in weakly ionised flows.  A detailed study of this
instability in the specific context of molecular clouds, and which is of
interest from the point of view of turbulence generation by stellar
outflows, is the subject of a future work.

In section \ref{sec:num-app} we outline the numerical and physical model
employed, in section \ref{sec:analysis} we discuss how we analyse the
results of our simulations while in section \ref{sec:results} we detail the 
results in both the linear and non-linear regimes, separating out the effects 
of ambipolar diffusion and the Hall effect in order to more fully understand 
the influence of each.

\section{Numerical approach}
\label{sec:num-app}

The simulations described in this work are performed using the HYDRA code 
\citep{osd06, osd07} for multifluid magnetohydrodynamics in the weakly 
ionised regime. We further assume the flow is isothermal.  The assumption of 
weak ionisation allows us to ignore the inertia of the charged species and 
allows us to derive a (relatively) straightforward generalised Ohm's law.  The 
resulting system of equations, given below, incorporates finite parallel, Hall and
Pederson conductivity and is valid in, for example, molecular clouds. In such regions the viscous lengthscales are much smaller than those over which nonideal effects are important. This leads to high Prandtl numbers and plasma flows in these regions can be considered to be effectively inviscid. In this 
work we examine low Mach number flows which, taken in concert with the 
isothermal assumption, means that features in the flow such as shocks are 
unlikely to create regions of high ionisation.

\subsection{Multifluid equations}

The HYDRA code solves the following equations for a system of $N$ fluids. The simulations described in this paper consist of three fluids, indexed by $i=0$ for the neutral fluid and $i=1$ and $i=2$ for the electron and ion fluids respectively. The equations to be solved are

\begin{equation}
\frac{\partial \rho_i}{\partial t} + \mathbf{\nabla} \cdot (\rho_i \mathbf{q}_i) = 0 \hspace{0.5cm} (0 \le i \le N-1),
\end{equation}

\begin{equation}
\frac{\partial \rho_0 \mathbf{q}_0}{\partial t} + \mathbf{\nabla} \cdot
	(\rho_0\mathbf{q}_0\mathbf{q}_0 + a^2\rho_0\mathbf{I}) = \mathbf{J} \times \mathbf{B},
\end{equation}

\begin{equation}
\frac{\partial \mathbf{B}}{\partial t} + \mathbf{\nabla} \cdot (\mathbf{q}_0 \mathbf{B} - \mathbf{B} \mathbf{q}_0) = - \mathbf{\nabla} \times \mathbf{E}^\prime,
\end{equation}

\begin{equation}
\alpha_i \rho_i (\mathbf{E} + \mathbf{q}_i \times \mathbf{B}) + \rho_i\rho_0K_{i0}(\mathbf{q}_0 - \mathbf{q}_i) = 0  \hspace{0.2cm} (1 \le i \le N-1),
\end{equation}

\begin{equation}
\mathbf{\nabla} \cdot \mathbf{B} = 0,
\end{equation}

\begin{equation}
\mathbf{\nabla} \times \mathbf{B} = \mathbf{J},
\end{equation}

\begin{equation}
\sum_{i=1}^{N-1}\alpha_i\rho_i = 0,
\end{equation}

\noindent where $\rho_i$, $\mathbf{q}_i$, $\mathbf{B}$, and $\mathbf{J}$ are the mass
	densities, velocities, magnetic field and current density, respectively. $a$ denotes the sound speed, and $\alpha_i$ and $K_{i0}$  are the charge-to-mass ratios and the collision coefficients between the charged species and the neutral fluid, respectively.

These equations lead to an expression for the electric field in the frame of the fluid, $\mathbf{E}^\prime$, given by the generalised Ohm's Law
\begin{equation}
\mathbf{E}^\prime = \mathbf{E}_{\rm O} + \mathbf{E}_{\rm H} + \mathbf{E}_{\rm A},
\end{equation}
where the components of the field are given by
\begin{equation}
\mathbf{E}_{\rm O} = (\mathbf{J} \cdot \mathbf{a}_{\rm O})\mathbf{a}_{\rm O},
\end{equation}

\begin{equation}
\mathbf{E}_{\rm H} = \mathbf{J} \times \mathbf{a}_{\rm H},
\end{equation}

\begin{equation}
\mathbf{E}_{\rm A} = -(\mathbf{J} \times \mathbf{a}_{\rm H}) \times \mathbf{a}_{\rm H},
\end{equation}

\noindent using the definitions $\mathbf{a}_{\rm O} \equiv f_{\rm O}\mathbf{B}$, $\mathbf{a}_{\rm H} \equiv f_{\rm H}\mathbf{B}$, $\mathbf{a}_{\rm A} \equiv f_{\rm A}\mathbf{B}$, where $f_{\rm O} \equiv \sqrt{r_{\rm O}}/B$, $f_{\rm H} \equiv r_{\rm H}/B$ and $f_{\rm A} \equiv \sqrt{r_{\rm A}}/B$. The resistivities given here are the Ohmic, Hall and ambipolar resistivities, respectively, and are defined by

\begin{equation}
r_{\rm O} \equiv \frac{1}{\sigma_{\rm O}},
\end{equation}

\begin{equation}
r_{\rm H} \equiv \frac{\sigma_{\rm H}}{\sigma_{\rm H}^2 + \sigma_{\rm A}^2},
\end{equation}

\begin{equation}
r_{\rm A} \equiv \frac{\sigma_{\rm A}}{\sigma_{\rm H}^2 + \sigma_{\rm A}^2},
\end{equation}
where the conductivities are given by

\begin{equation}
\sigma_{\rm O} = \frac{1}{B} \sum_{i=1}^{N-1} \alpha_i\rho_i\beta_i,
\end{equation}

\begin{equation}
\sigma_{\rm H} = \frac{1}{B} \sum_{i=1}^{N-1} \frac{ \alpha_i\rho_i}{1+\beta_i^2},
\end{equation}

\begin{equation}
\sigma_{\rm A} = \frac{1}{B} \sum_{i=1}^{N-1} \frac{ \alpha_i\rho_i\beta_i}{1+\beta_i^2},
\end{equation}
where the Hall parameter $\beta_i$ for a charged species is given by 
\begin{equation}
\beta_i = \frac{\alpha_i B}{K_{i0}\rho_0} .
\end{equation}

To solve these equations numerically we use three different operators:
\begin{enumerate}
\item solve equations (1), (2), (3), including the restriction of
	equation (5) and for $i=0$, using a standard second order, finite volume 
	shock-capturing scheme.  Note that for this operator the resistivity terms 
	in equation (3) are not incorporated.  Equation (5) is incorporated using 
	the method of Dedner \citep{dedner}.
\item Incorporate the resistive effects in equation (3) using 
super-time-stepping to accelerate the ambipolar diffusion term and the
Hall Diffusion Scheme to deal with the Hall term.
\item Solve equations (4) for the charged species velocities and use
these to update equation (1) (with $i=1,\ldots,N-1$)
\end{enumerate}
These operators are applied using Strang operator splitting in order to
maintain the second order accuracy of the overall scheme.  We refer the
reader to \cite{osd06, osd07} for a more detailed description.

\subsection{Initial conditions}
% Description + diagram of set-up + table and nomenclature of simulations?
\label{subsec:init-cond}

The simulations are carried out on a 2.5\,D slab grid in the $xy$-plane.
The grid consists of $6400 \times 200 \times 1$ cells, in the $x$, $y$,
and $z$ directions respectively. This resolution was chosen on the basis
that it reproduces the initial linear growth of the ideal MHD system in 
\cite{keppens99}.  Resolution studies were performed to confirm the
resolution as being appropriate (see Sect.\ \ref{sec:ambi-res-study} and
\ref{sec:hall-res-study}).

The initial set-up used was that of two plasmas flowing anti-parallel
side-by-side on a grid of size $x = [0, 32L]$ and $y = [0, L]$. The
plasma velocities are given by $+ \frac{V_0}{2}$ and $-\frac{V_0}{2}$ in
the $y$-direction, with a tangential shear layer of width $2a$ at the
interface at $x = 16L$. This velocity profile is described by
\begin{equation}
\mathbf{v}_0 =  \frac{V_0}{2} \tanh \left( \frac{x - 16L}{a} \right)
	\hat{\mathbf{y}} .
\end{equation}
\noindent The width of the shear layer is chosen to be $\frac{a}{L} =
0.05$, or approximately 20 grid zones.  The magnetic field is initially
set to be uniform and aligned with the plasma flow.  

The initial background for all three fluids in the system is now an exact 
equilibrium.
The initial neutral velocity field, $V_0$ is then augmented with a perturbation
given by
\begin{equation}
\delta v_x = \delta V_0 \sin( - k_y y) \exp \left( - \frac{(x -
			16L)^2}{\sigma^2} \right).
\end{equation}
where $\delta V_0$ is set to $10^{-4}\,V_0$.  The wavelength of the
perturbation is set equal to the characteristic length scale, $\lambda
= \frac{2 \pi}{k_y} = L$, so that a single wavelength fits exactly into
the computational domain. This maximises the possibility of resolving
structures that are small relative to the initial perturbed wavelength,
\citep{frank96}. The perturbation attenuation scale is chosen so that it is 
larger than the shear layer, but small enough so that the instability can be 
assumed to interact only minimally with the $x$-boundaries 
\citep[see][]{palotti08}, and is set using $\frac{\sigma}{L} = 0.2$ 
\citep[see][]{keppens99, palotti08}.

The physical parameters are then chosen using normalised, dimensionless
quantities. The wavenumber $k_y$ is chosen to be $2 \pi$ in order to
maximise the growth rate of the instability \citep{keppens99}. This
normalises the length scale of the simulation so that $L = 1$. The
timescale is then normalised by the sound speed, so that $c_{\rm s} =
\frac{L}{T} = 1$. The mass scale is chosen such that the initial mass
density is set to unity, $\rho_0 = 1$. In the isothermal case the
adiabatic index $\gamma = 1$, and this gives us a sound speed $c_{\rm s}
\equiv \sqrt{\frac{\gamma p_0}{\rho_0}} = 1$, and the initial pressure
is therefore also equal to unity, $p_0 = 1$. A transonic flow is chosen
with sonic Mach number $M_{\rm s} = \frac{V_0}{c_{\rm s}} = 1$, so that
the plasma has velocity $\pm \frac{V_0}{2} = \pm 0.5$. In order for the
KH instability to grow, it is required that the flow be super-Alfv\'{e}nic 
\citep{chand61}, and so for this study, an Alfv\'{e}n Mach number
$M_{\rm A} = \frac{V_0}{v_{\rm A}} = 10$ is chosen. This sets the
Alfv\'{e}n velocity, $v_{\rm A} = \sqrt{\frac{B_0^2}{\rho_0}} = 0.1$,
and the initial magnetic field strength to $B_0 = 0.1$.  

We use periodic boundary conditions at the high and low $y$ boundaries.  Since 
we wish to study not only the initial growth phase of the instability, but 
also its subsequent non-linear behaviour we must ensure that waves interacting
with the high and low $x$ boundaries do not reflect back into the domain
to influence the dynamics.  Several test simulations for various
parameters have shown that a large width of $32$ is necessary to
ensure this.  We use gradient zero boundary conditions at the high and
low $x$ and $z$ boundaries.  

Finally we must choose the parameters describing the properties of our
charged fluids in our multifluid system.  Our basic parameter set is
contained in table \ref{table:mf-params}.  See \cite{jones11} for more
details.

\begin{table}
\begin{tabular}{cccc} \hline
Fluid & Density & $K_{i0}$ & $\alpha_i$ \\ \hline
1 & $2.84\times10^{-13}$ & $2\times10^4$ & $-1\times10^{17}$ \\
2 & $2\times10^{-7}$ & $1.42\times10^{10}$ & $1.42\times10^{11}$ \\ \hline
\end{tabular}
\caption{\label{table:mf-params} The density, collision coefficient
($K_{i0}$) and charge-to-mass ratios ($\alpha_i$) for each of the
charged fluids in simulation full-low-hr (see table 
\ref{table:nomenclature}).  These parameters are modified to vary the 
resistivities as necessary for the other simulations in this work.  See text.}
\end{table}

Table \ref{table:nomenclature} contains the nomenclature we 
will use for the rest of this paper when referring to the simulations.
Each simulation is denoted by xxxx-yyyy-zz where the first set of
characters denote the dominant resistivity (Hall or ambipolar or
``full'' if both resistivities have the same magnitudes),
the second set denote the level of the resistivity (low, medium or high) and 
the final two characters denote the resolution (low, medium or high resolution 
denoted by lr, mr and hr respectively).  The high resolution 
simulations ($6400 \times 200$) took of the order of 1500 -- 3000 core hours 
on a quad-core Xeon E5430 based system.

\begin{table}
\begin{tabular}{cccc}\hline
Simulation & Resolution & $r_{\rm H}$ & $r_{\rm A}$ \\ \hline
mhd-zero-hr & $400\times200$ & 0 & 0 \\
hd-zero-hr & $400\times200$ & 0 & 0 \\
ambi-high-lr & $1600\times50$ & $3.52\times10^{-6}$ & $3.5\times10^{-2}$ \\
ambi-high-mr & $3200\times100$ & $3.52\times10^{-6}$ & $3.5\times10^{-2}$ \\
ambi-high-hr & $6400\times200$ & $3.52\times10^{-6}$ & $3.5\times10^{-2}$ \\
full-low-hr & $6400\times200$ & $3.52\times10^{-6}$ & $3.5\times10^{-6}$ \\
ambi-med-hr & $6400\times200$ & $3.52\times10^{-6}$ & $3.5\times10^{-3}$ \\
hall-high-lr & $1600\times50$ & $3.52\times10^{-2}$ & $3.5\times10^{-6}$ \\
hall-high-mr & $3200\times100$ & $3.52\times10^{-3}$ & $3.5\times10^{-6}$ \\
hall-high-hr & $6400\times200$ & $3.52\times10^{-2}$ & $3.5\times10^{-6}$ \\
hall-med-hr & $6400\times200$ & $3.52\times10^{-3}$ & $3.5\times10^{-6}$ \\
\\ \hline
\end{tabular}
\caption{\label{table:nomenclature} The resolutions and (initial) 
resistivities of each of the simulations presented in this work. The
Hall resistivity is denoted by $r_{\rm H}$ while the ambipolar
resistivity is denoted by $r_{\rm A}$. Note that the Pederson
conductivity is always very large and hence its associated resistivity
is always several orders of magnitude smaller than both $r_{\rm H}$ and
$r_{\rm A}$.}
\end{table}

Note that simulation full-low-hr is effectively an ideal MHD simulation
and produced results virtually identical to mhd-zero-hr which is a true
MHD simulation (see Sect.\ \ref{sec:validation})and which was used for
comparison with previous literature. %and also to determine the influence of the non-ideal terms in the rest of our suite of simulations.

\section{Analysis}
\label{sec:analysis}

In order to study the growth of the instability, the evolution of a
number of parameters can be measured with time. In particular, we
measure the transverse kinetic energy $$\Ek{x} \equiv \int \! \int 
\frac{1}{2} \rho v_x^2\,dx\,dy$$ and the magnetic energy $$E_b \equiv
\frac{1}{2} \int \! \int \left\{\left[ B_x^2 + B_y^2 + B_z^2
\right] - B_0^2\right\}\,dx\,dy$$ in the
system where $B_0$ is the magnitude of the magnetic field at $t=0$. 
As the entire plasma flow is initially in the $y$-direction, with only a very 
small perturbation in the $x$-direction, any growth of $\Ek{x}$ is
due to the growth of the instability.  It is possible to determine the
growth rate of the instability directly from the growth of the
transverse kinetic energy \citep{keppens99}: the transverse kinetic can be 
expressed as
\begin{eqnarray}
\Ek{x} &=& (\rho_0 + \delta \rho)(|v_x|_0 + \delta v_x)^2 \\
     &=& (\rho_0 + \delta \rho)(\delta v_x)^2 \\
	  &\approx& \rho_0  \delta v_x^2 \\
     & \propto & \exp[2i(\mathbf{k} \cdot \mathbf{r} - \omega t)]
\end{eqnarray}
presuming the initial perturbation is proportional to
$\exp[i(\mathbf{k} \cdot \mathbf{r} - \omega t)]$.  Hence the kinetic
energy grows at a rate of $2 \omega$, where $\omega$ is the growth rate of the 
instability.

\section{Results}
\label{sec:results}

We start by describing the validation of the set-up used by comparing an
ideal MHD simulation run using HYDRA with previously 
published literature.  We then go on to discuss the behaviour of the KH
instability in ambipolar-dominated and Hall-dominated flows
respectively.  

\subsection{Validation}
\label{sec:validation}

\begin{figure}
%
% Aoife's figure chapt3_linear_logKEx2.png
%
\centering
\includegraphics[width=8.4cm]{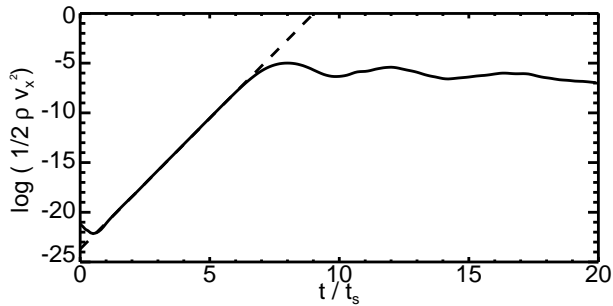}
\caption{Plot of the log of the transverse kinetic energy with time for
mhd-zero-hr. A linear function (dashed line) with slope 2.63 has been 
over-plotted.} 
 \label{fig:ideal_validation_ek}
\end{figure}

In order first to validate our set-up we examine our mhd-zero-hr simulation 
with HYDRA and determine the growth rate using the kinetic energy of motions
in the $x$ direction as described in Sect.\ \ref{sec:analysis}.  
%Note that the magnetic Reynolds number, $Re_{\rm m}$, for this simulation is formally $2.84\times10^5$ so this is effectively an ideal MHD simulation. 
Figure \ref{fig:ideal_validation_ek} contains a plot of the log of the transverse 
kinetic energy as a function of time.  At early times this growth is clearly
exponential and can be fitted with a line of slope 2.63 implying a
growth rate, normalised by the width of the shear layer and the
initial relative velocity, for the dominant mode of the KH instability of 
0.1315.  We can compare this with the value of the growth rate calculated
analytically by \citet{miura82} (their Figure 4) at this wavenumber of
0.13.  While comparisons between linear studies of incompressible flows,
and numerical studies of compressible flows are bound to differ to some 
extent, these results are seen to agree exceptionally well.

\begin{figure}
%
% Aoife's figure chapt3_linear_total_magE2.png
%
\centering
\includegraphics[width=8.4cm]{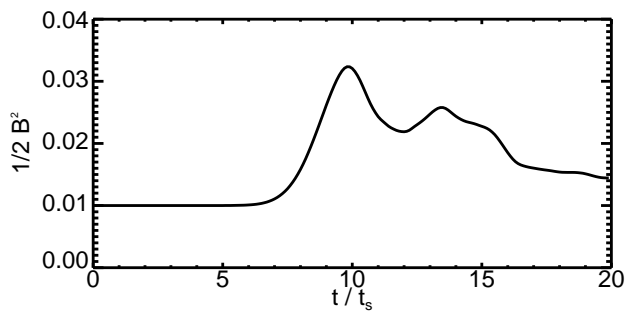}
\caption{Plot of the evolution of magnetic energy with time for
simulation mhd-zero-hr.} 
\label{fig:ideal_validation_b}
\end{figure}

We wish to examine not only the linear regime but also the non-linear
regime.  We compare our results for the growth of magnetic energy with
those of \citet{mala96} (the upper panel of their Figure 5).  Figure
\ref{fig:ideal_validation_b} contains a plot of the magnetic energy,
calculated as $\int \! \int  \frac{1}{2} \big( B_x^2 + B_y^2 + B_z^2
\big)\,dx\,dy $, as a function of time.  The maximum magnetic energy
reached in our simulations matches that of \citet{mala96} to within
10\%.  Our simulation reaches saturation at a later time but the exact
time of saturation depends on the initial amplitude of the perturbation
and so this is not a concern.

We are therefore confident of the behaviour of HYDRA in simulating the
KH instability.  We now move on to investigating the influence of
multifluid MHD effects on the growth, saturation and non-linear
behaviour of this instability.

\subsection{Ambipolar dominated flows}
\label{sec:ambi-flows}

We begin our study of the multifluid KH instability by choosing our fluid 
parameters to ensure our ambipolar resistivity is dynamically
significant while minimising the Hall resistivity.  This allows us 
to isolate the influence of the ambipolar resistivity on the instability.  In 
order to increase the
ambipolar resistivity we change the value of the collision coefficient
for species 2 so that $K_{2,0}$ is decreased by 3 orders of
magnitude from that given in table \ref{table:mf-params} for simulation
ambi-med-hr and by 4 orders of magnitude for ambi-high-hr.  These alterations 
of $K_{2,0}$ give values of $r_{\rm A}$ of $3.5\times10^{-3}$ and 
$3.5\times10^{-2}$ respectively, and (ambipolar) magnetic Reynold's numbers, $Re_{\rm m}$, of 
$2.84\times10^2$ and $2.84 \times 10^1$ respectively. 
These simulations are examined in comparison to the full-low-hr simulation. With a formal magnetic Reynolds number $2.84\times10^5$, the diffusion in this set-up is predominantly numerical, and as such, it is effectively an ideal MHD simulation.
	
\subsubsection{Resolution study}
\label{sec:ambi-res-study}

In non-ideal MHD we must ensure that the length scales over which the
diffusion of the magnetic field (or the whistler waves in the case of
Hall dominated flows) must be resolved in order to properly track the
dynamics of the system.  To this end we perform a resolution study using
simulations ambi-high-lr, ambi-high-mr and ambi-high-hr (see Table
\ref{table:nomenclature}). 

\begin{figure}
\centering
\includegraphics[width=8.4cm]{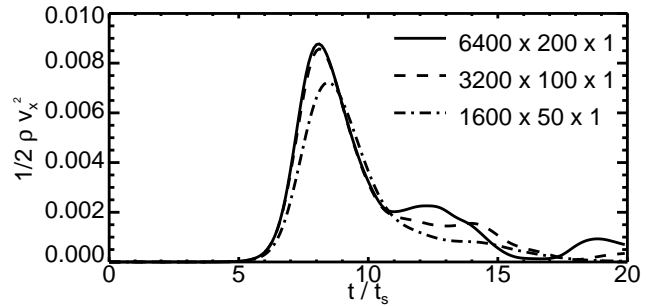}
\caption{Plot of the evolution of the transverse kinetic energy against
time in the case of high ambipolar resistivity for ambi-high-lr,
ambi-high-mr and ambi-high-hr.} 
\label{fig:ambi-res-ek}
\end{figure}

\begin{figure}
\centering
\includegraphics[width=8.4cm]{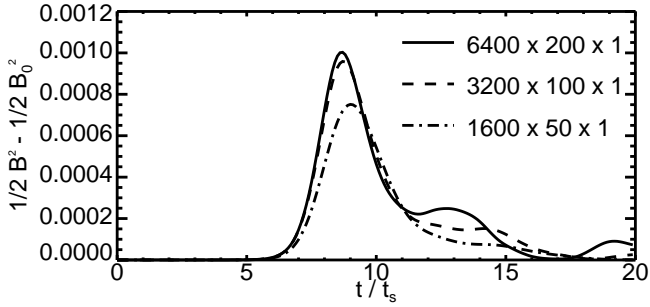}
\caption{As figure \ref{fig:ambi-res-ek} but for the perturbed magnetic energy.}  
\label{fig:ambi-res-eb}
\end{figure}

Figures \ref{fig:ambi-res-ek} and \ref{fig:ambi-res-eb} contain plots of the
evolution of $\Ek{x}$ and $E_{\rm b}$ for each of the simulations in our
resolution study.  It can be seen that the linear growth in ambi-high-lr is
significantly lower than the two other simulations.  However, the linear
behaviour is almost identical for ambi-high-mr and ambi-high-hr.  The 
subsequent non-linear behaviour is similar with only relatively small
variations after $t \sim 11$.

The results of this study indicate that a resolution of $6400\times200$
is sufficient to capture the initial growth and saturation of the instability.  
Subsequently, the dynamics is captured at least qualitatively.

%
% Figure fig:ambi-res-ek is 4.8 in Aoife's thesis
% Figure fig:ambi-res-eb is 4.9 in Aoife's thesis
%

\subsubsection{The linear regime}
\label{sec:ambi-linear}
%
%  Demonstrate the similarity between ideal and ambi linear growth.
%  Refer to future sections which will also show the same thing.
%

\begin{figure}
\centering
\includegraphics[width=8.4cm]{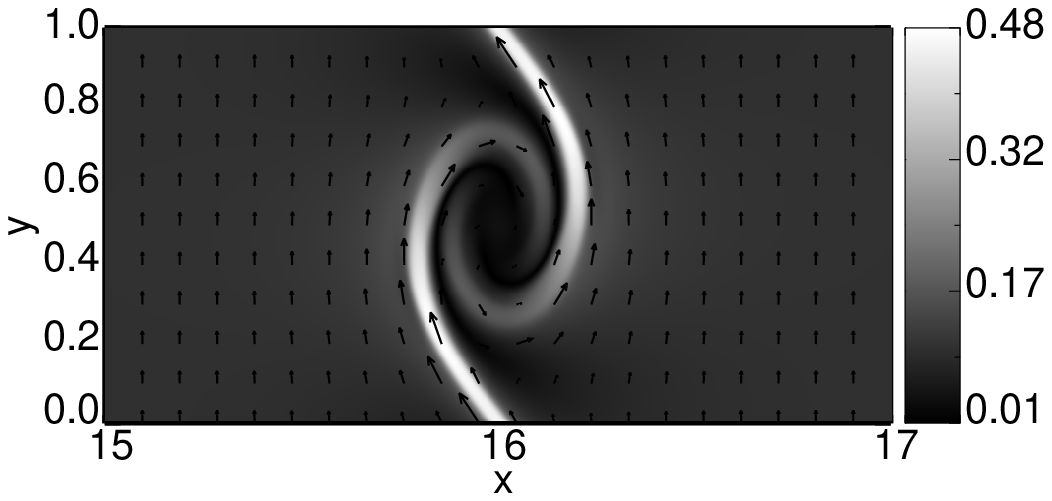}
\vspace{0.1cm}
\includegraphics[width=8.4cm]{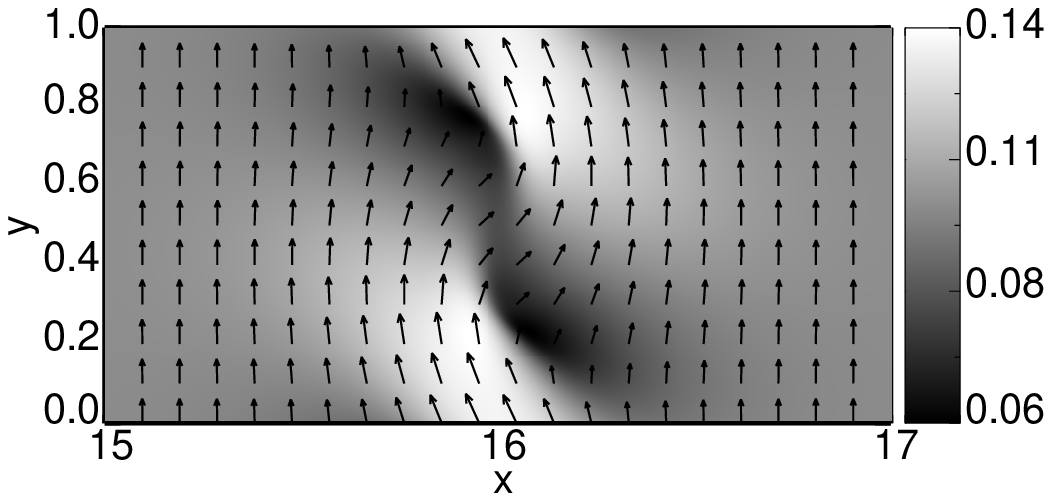}
\caption{Plots of the magnitude and vector field of the magnetic field
for full-low-hr, at time $t = 8 \, t_{\rm s}$ (upper panel) and for
ambi-high-hr (lower panel).  The magnetic field is wound up in a similar
fashion to the velocity in the upper panel, but not the lower panel.} 
\label{fig:b-field-comp}
\end{figure}

Generally speaking, the evolution of the KH instability in ideal MHD
leads to a wind-up of both the plasma and the magnetic field at the
interface between the two fluids, resulting in the ``Kelvin's cat's
eye'' vortex.  Multifluid effects alter the nature of this significantly.  
Figure \ref{fig:b-field-comp} contains plots of the magnetic field at 
$t=8$\,$t_{\rm s}$ which is the time at which the instability saturates for 
both full-low-hr (effectively ideal MHD) and ambi-high-hr.  The difference in 
the morphology is clear.  

\begin{figure}
\centering
\includegraphics[width=8.4cm]{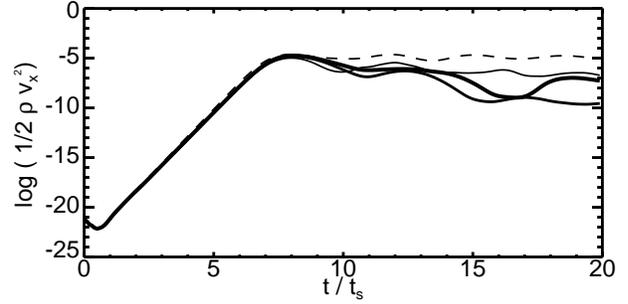}
\caption{Plots of the log of the transverse kinetic energy for varying 
ambipolar resistivity. The thicker lines represent the models with higher 
ambipolar resistivity. The dashed line represents hd-zero-hr. There is very 
little difference between the linear growths for the three magnetised cases.} 
\label{fig:ambi-ek}
\end{figure}

\begin{figure}
\centering
\includegraphics[width=8.4cm]{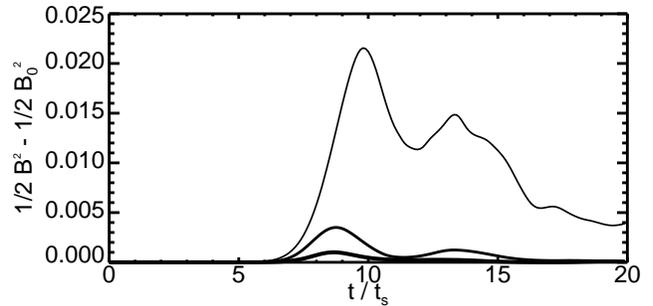}
\caption{The perturbed magnetic energy in the system is plotted against time 
in the three cases of varying ambipolar resistivity. The thicker lines 
represent the models with higher ambipolar resistivity. Higher ambipolar 
resistivity causes the magnetic field to experience less amplification, 
through diffusion and decoupling of the bulk fluid from the 
magnetic field.} 
\label{fig:ambi-eb}
\end{figure}

Given these striking differences at saturation it is interesting to
investigate whether the linear growth rate is influenced by the addition
of ambipolar diffusion.  Figure \ref{fig:ambi-ek} contains plots of the
evolution of $\Ek{x}$ with time for the various simulations.  The
linear growth rate remains almost unchanged with the addition of
ambipolar diffusion.  On the other hand, figure \ref{fig:ambi-eb}
contains plots of $E_b$ as a function of time.  It is clear that
the perturbed magnetic energy is strongly influenced, even well before
saturation, by the presence of ambipolar diffusion.  We will discuss
this in more detail in section \ref{sec:ambi-nonlinear}.  

%
%  Figure fig:b-field-comp is a concatenation of figures 4.14 and 4.15
%  in Aoife's thesis.
%
%  Figure fig:ambi-ek is figure 4.16 in Aoife's thesis (or 4.26)
%  Figure fig:ambi-eb is figure 4.13 in Aoife's thesis
%  

%
%  Do I have refs for previous studies with linear ambipolar stuff?
%  What about a short paper with Sarah on the linear behaviour of the
%  multifluid equations?
%

\subsubsection{The nonlinear regime}
\label{sec:ambi-nonlinear}
%
%  - Introduction of reconnection (plot of B energy)
%

It is clear from figure \ref{fig:ambi-eb} that the magnetic energy is
strongly influenced by the presence of ambipolar diffusion.  This is not
too surprising as ambipolar diffusion, being a genuinely diffusive
process (unlike the Hall effect) allows the magnetic field to diffuse relative to the bulk flow.
%reconnection which is a path through which the system can dispose of energy.  

As the collision rate between the ion and neutral fluids is decreased, the ion 
fluid decouples from the bulk fluid, and thus the magnetic field 
becomes decoupled from the bulk flow: the frozen-in approximation of ideal 
MHD is broken. 
%The topology of the field need therefore no longer be preserved and magnetic reconnection can occur. 
As a result, the magnetic field is able to diffuse through the bulk fluid rather than being tied to it.
In figure \ref{fig:ambi-eb}, 
%magnetic reconnection 
diffusion can be identified as the cause of the decrease of the amplification of the magnetic energy with time for increasing ambipolar resistivity. A cursory examination of the topology of the magnetic
field for the low resistivity case (full-low-hr in figure 
\ref{fig:b-field-comp}) demonstrates clearly that there are regions in the 
flow which will be susceptible to diffusion%reconnection
: regions in which the magnetic field lines have been compressed and amplified. 
%anti-parallel magnetic field lines are in close proximity.

It is clear, again from figure \ref{fig:ambi-eb}, that there
	is a significant decrease in the growth
%amount 
of magnetic energy 
%is lost through reconnection 
as a result of diffusion for even a
moderate amount of ambipolar resistivity.  
%This reconnection 
The diffusion has an
influence on the dynamics of the neutral fluid: 
%once the magnetic energy is lost through reconnection 
when the magnetic energy has not been as strongly amplified, the field can no longer exert the
same effect on the neutral and ion fluid.  Examination of figure 
\ref{fig:ambi-ek} reveals that the peak reached by $\Ek{x}$ {\em
increases} with increasing ambipolar resistivity. The value reached tends
toward the hydrodynamic limit for two reasons: increasing resistivity
implies increasing 
%reconnection 
diffusion and hence a decreasing field strength,
and it also implies less coupling between the magnetic field and the neutral 
fluid with increasing resistivity. As expected, the weaker magnetic field strength allows the vortex to become more rolled up \citep{faganello}.

\begin{figure}
\centering
\includegraphics[width=8.4cm]{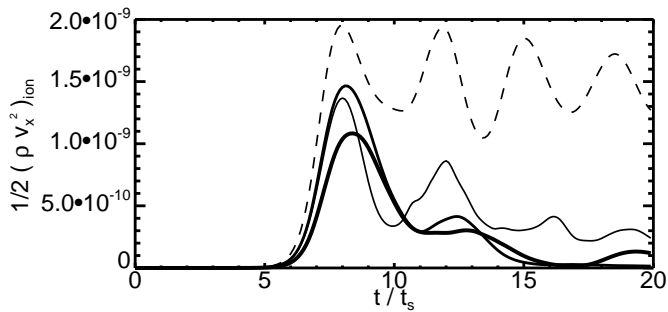}
\caption{Plots of the transverse kinetic energy of the ion fluid as a
function of time for each of full-low-hr, ambi-med-hr, ambi-high-hr and 
hd-zero-hr. The thicker lines represent the models with higher ambipolar 
resistivity. The dashed line represents the hydrodynamic case.} 
\label{fig:ambi-ek-ion}
\end{figure}

It can be shown that for the ambi-med-hr simulation, the increase in the 
wind-up of the bulk fluid is due solely to the first source: the magnetic diffusion.
%reconnection. 
The slight decoupling of the ion and neutral fluid is 
sufficiently high to allow magnetic %reconnection
diffusion, while still being 
sufficiently low to force the charged fluids to behave in a manner similar to 
the bulk fluid. This can be seen in a plot of the transverse kinetic energy of 
the ion fluid, as shown in figure \ref{fig:ambi-ek-ion}.  With moderate
amounts of ambipolar diffusion, the transverse kinetic energy of the ion 
fluid (and electron fluid) reaches a higher maximum, meaning that the 
fluid experiences more wind-up.

However, as the ambipolar resistivity is increased further, the ion fluid 
becomes more decoupled from the neutral fluid, and is instead influenced 
primarily by the dynamics of the magnetic field. This significant 
decoupling occurs only for magnetic Reynolds number lower than $Re_{\rm
m} \approx 100$. As the magnetic field therefore undergoes less wind-up than 
before, so does the ion (and electron) fluid. This decrease in the winding-up 
of the charged fluid velocity field is the cause of the decrease in the 
transverse kinetic energy of the ion fluid, as is seen in
ambi-high-hr in figure \ref{fig:ambi-ek-ion}.

%
%  Figure ambi-ek-ion is figure 4.17 in Aoife's thesis
%

%  - Two sources of B energy loss/lack of energy growth: reconnection, 
%	                                                      decoupling
%  - Increase in transverse (neutral) kinetic energy
%  - Geometry of flow - vortext break-up or not etc.
\subsection{Hall dominated flows}
\label{sec:hall-flows}
%
% Description of how to make the flow Hall dominated (i.e. the K's and
% the alpha's).  
%
We now turn our attention to the likely influence of the Hall effect on
the KH instability in multifluid MHD flows.  In order to attain the 
resistivities we want (see table
\ref{table:nomenclature}) we reduce the charge-to-mass ratio of the ion
fluid, causing its Hall parameter to become smaller, causing the ion
fluid dynamics to more closely resemble the bulk (neutral) fluid
dynamics.  The electrons are, however, still well-tied to the field
lines and so a relative drift emerges between the ion and electron
fluids leading to a current perpendicular to the magnetic field and
hence the Hall effect.

\subsubsection{Resolution study}
\label{sec:hall-res-study}
%
% Note the necessary width of the computational domain and give some
% energy plots to demonstrate we're doing OK on the resolution front.
%

As in the ambipolar resistivity study, a resolution study is carried out
to ensure that the small-scale non-ideal dynamics are captured. For this 
purpose, simulations are again run at three different resolutions (see
Table \ref{table:nomenclature}). These simulations are run with the highest 
level of Hall resistivity to ensure that the smallest-scale dispersive effects 
are in place when examining whether they are sufficiently resolved. 
The inclusion of the Hall term in the dispersion relation in
principle allows for the introduction of waves with a signal speed which tends 
towards infinity as their wavelength tends towards zero. While the Hall term 
is handled in the equations by the HYDRA code using the explicit Hall 
Diffusion Scheme (HDS) \citep{osd06, osd07}, the code naturally does not 
resolve these waves of vanishing wavelength.

\begin{figure}
\centering
\includegraphics[width=8.4cm]{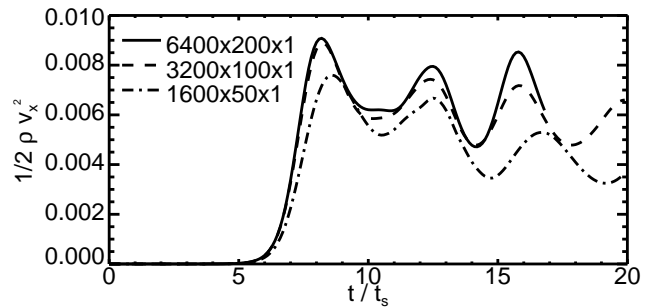}
\caption{Plot of the transverse kinetic energy against time for
simulations hall-high-lr, hall-high-mr and hall-high-hr.} 
\label{fig:hall-res-ek}
\end{figure}

As has been seen, the introduction of non-ideal effects does not tend to 
greatly influence the linear growth rate of the instability. As a result, 
the growth rate does not provide a good means of measuring convergence 
with increasing resolution.  The nonlinear evolution of the transverse
kinetic energy, $\Ek{x}$ is, however, strongly influenced by the
non-ideal effects.  We do not examine the evolution of the 
perturbed magnetic energy as, in the case of high Hall resistivity, it no 
longer demonstrates a simple growth to an initial maximum.  The evolution of 
the transverse kinetic energy for each simulation is plotted in figure
\ref{fig:hall-res-ek}.  It can clearly be seen that the simulations have 
started to converge at higher resolutions.  While there is a notable 
difference between the two simulations of lower resolution, the gap closes 
significantly in the comparison between the two simulations of higher 
resolution.  In particular, the initial maxima of $\Ek{x}$ are almost 
identical in simulations hall-high-mr and hall-high-hr.

We are, therefore, confident that the dynamics in the hall-hr are well 
resolved and that our conclusions as to the physical processes occurring are 
well-founded.
%
%  fig:hall-res-ek is figure 5.1 in Aoife's thesis

\subsubsection{The linear regime}
\label{sec:hall-linear}
%
%  Heavily reference {sec:ambi-linear} as we have the same result here:
%  that the linear regime is largely the same as in the idea/low
%  resistivity case.  Growth of Bz also.
%

\begin{figure}
\centering
\includegraphics[width=8.4cm]{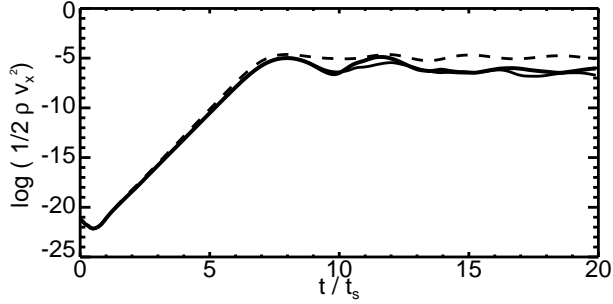}
\caption{The log of the transverse kinetic energy in the system is plotted 
against time for the full-low-hr (thin line) and hall-med-hr (thick line). The 
dashed line represents hd-zero-hr. There is very little difference between the 
linear growths for the magnetised systems.} 
\label{fig:hall-ek}
\end{figure}

\begin{figure}
\centering
\includegraphics[width=8.4cm]{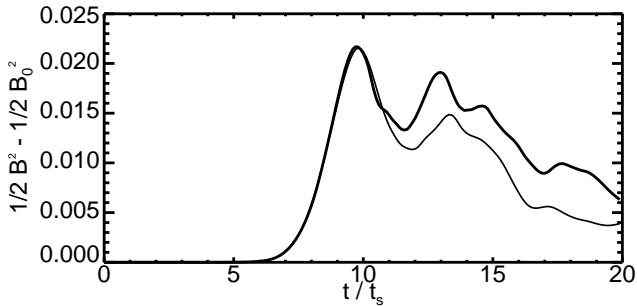}
\caption{The perturbed magnetic energy in the system for both
full-low-hr (thin line) and hall-med-hr (thick line) are plotted against time. 
No significant difference can be seen between the maxima reached in each 
case.} 
\label{fig:hall-eb}
\end{figure}

Figures \ref{fig:hall-ek} and \ref{fig:hall-eb} contain plots of $\Ek{x}$ and 
$E_b$ as a function of time for simulations hall-med-hr and
full-low-hr.  In the linear regime, the influence of the Hall effect on the 
growth rate of the instability, and on the magnetic and kinetic energy in the 
system at saturation can be seen to be negligible.  Hence we can conclude that 
neither ambipolar diffusion nor the Hall effect influence the energetics of 
the system in the linear regime.

\begin{figure}
\centering
\includegraphics[width=8.4cm]{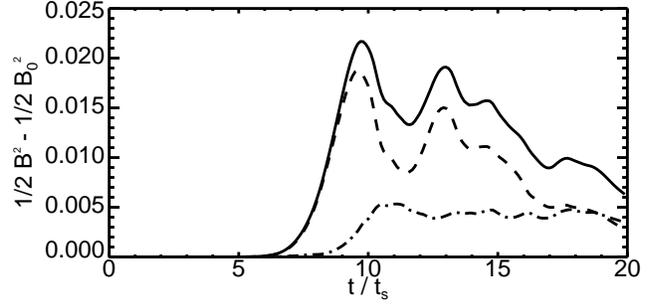}
\caption{The total perturbed magnetic energy (solid line) in the system is 
plotted against time for hall-med-hr. Also plotted for this simulation are the 
evolutions of the perturbed magnetic energies in the $xy$-plane only (dashed 
line) and in the $z$-direction (dot-dashed line). While the maximum reached by 
the total magnetic energy is the same as in full-low-hr, the magnetic field has 
experienced some re-orientation into the $z$-direction.} 
\label{fig:hall-eb-xy-z}
\end{figure}

We expect the Hall effect to re-orient the magnetic field out of the 
$xy$-plane in which it resides at $t=0$.  Figure \ref{fig:hall-eb-xy-z} 
contains plots of $E_b$, the 
perturbed magnetic energy in the $xy$-plane and the perturbed magnetic energy 
in the $z$ direction as functions of time.  It is clear that there is some 
growth of the magnetic field in the $z$ direction.  Interestingly, at 
saturation the magnetic energy in the $xy$-plane is noticeably less than 
$E_b$ and yet the overall value of $E_b$ is the same as that 
derived from the full-low-hr (i.e.\ quasi-ideal MHD) simulation.  An 
interesting point to note about this is that, whereas it has been shown
\citep{jones97} that the strength of the field {\em in the direction of the 
initial flow} is what is important in determining the effect of the magnetic 
field on the KH instability, here we can see that the strength of the field 
perpendicular to this plane appears to play a role also.

%
%  NB: REALLY NEED TO CHECK THIS: read original papers on it
%

%
%
%

\subsubsection{The nonlinear regime}
\label{sec:hall-nonlinear}
%
%  Growth of Bz, swaps between E_B and E_k (amplitudes of oscillations
%  are larger than in low res case), growth of kinetic energy in
%  the z direction.

\begin{figure}
\centering
\includegraphics[width=8.4cm]{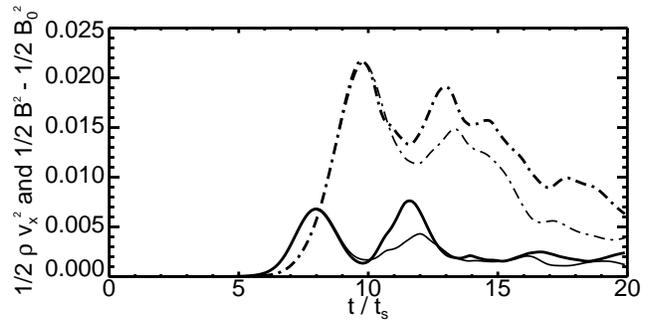}
\caption{Plot of the total perturbed magnetic energy (dot-dashed lines) and 
transverse kinetic energy (solid lines) with time for the full-low-hr 
(thin lines) and hall-med-hr (thick lines) simulations.} 
\label{fig:eb-ek}
\end{figure}

Following the initial linear growth of the instability, the system
experiences a period of transferring energy back and forth between the
magnetic field and velocity field, as in full-low-hr (see figure 
\ref{fig:eb-ek}).  Interestingly, the amplitude of the oscillations - i.e.\ 
the amount of energy being transferred between motion and the magnetic field - 
is larger in hall-med-hr than in full-low-hr.  This may be due to 
re-orientation of the magnetic field out of the plane of the instability and 
hence a reduction in (numerical) reconnection.  It is worth recalling here 
that the Hall effect, although it appears similar to a diffusion term in the 
induction equation, is a dispersive effect which conserves magnetic energy.

Perhaps the most important difference between full-low-hr and
hall-med-hr is that while the kinetic energy in the $y$-direction
gradually tends towards a constant value in the low
resistivity case, it continues to steadily decay in hall-med-hr. The 
conclusion is that in full-low-hr, the instability has completed its growth 
and is returning to a quasi-steady state. In hall-med-hr however, the 
instability is continuing to consume the parallel kinetic energy available to 
it and the instability undergoes a further stage of development as a
result of the inclusion of the Hall effect.

%
% Figure fig:eb-ek is 5.11 in ACJ
%

\begin{figure}
\centering
\includegraphics[width=8.4cm]{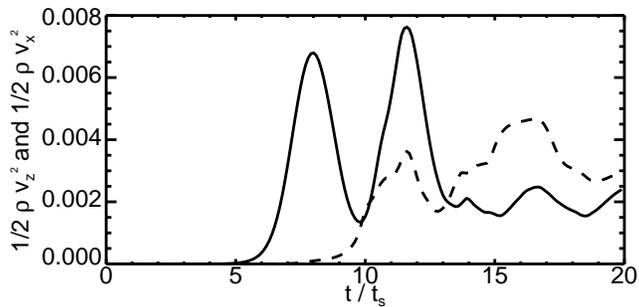}
\caption{Plot of the transverse kinetic energy (solid line) and kinetic energy 
in the $z$-direction (dashed line) with time for hall-med-hr. It can be 
seen that at later times, the growth of energy in the $z$-direction has become 
significant.} 
\label{fig:ek-ekz}
\end{figure}

\begin{figure}
\centering
\includegraphics[width=8.4cm]{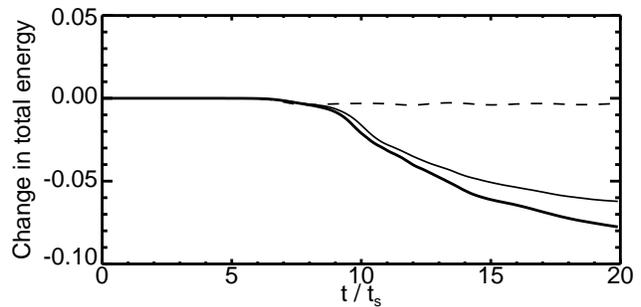}
\vspace{0.1cm}
\includegraphics[width=8.4cm]{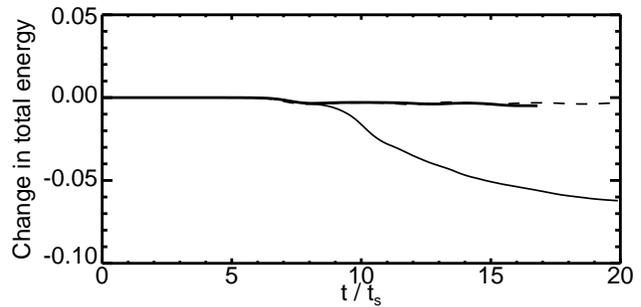}
\caption{The change in total energy of the system is plotted against time for 
hall-med-hr (thick line, upper panel) and hall-high-hr cases (thick line, 
lower panel).  In each case the change in total energy is plotted for
full-low-hr (thin line) for comparison and also for hd-zero-hr (dashed
line).} 
\label{fig:total-e}
\end{figure}

\begin{figure}
\centering
\includegraphics[width=8.4cm]{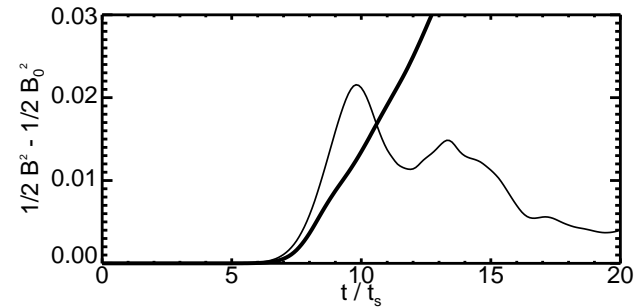}
\caption{The total perturbed magnetic energy in the system is plotted against 
time for full-low-hr (thin line) and hall-high-hr (thick line) cases.  The 
magnetic energy is seen to experience very different growth in
hall-high-hr.} 
\label{fig:eb-hall}
\end{figure}

\begin{figure}
\centering
\includegraphics[width=8.4cm]{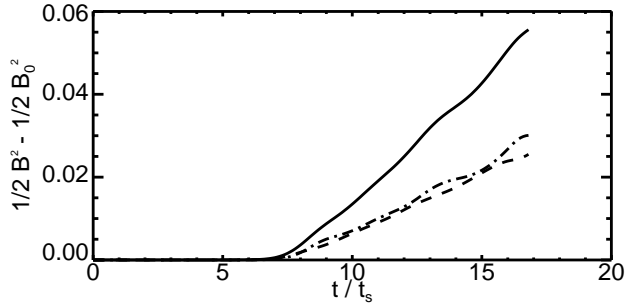}
\caption{The perturbed magnetic energy in the system is plotted against time 
for hall-high-hr (solid line). Also plotted for comparison are the growth of 
the magnetic energies in the $xy$-plane (dashed line) and in the 
$z$-direction (dot-dashed line). It can be seen that there is significant 
growth in the $z$-direction.} 
\label{fig:eb-hall-high-z}
\end{figure}

Since the magnetic field gains a component in the $z$ direction in
hall-med-hr due to the Hall effect it is interesting to examine the behaviour 
of the kinetic energy in the $z$ direction also.  Figure \ref{fig:ek-ekz}
contains plots of the transverse kinetic energy and the kinetic energy
in the $z$ direction (i.e.\ $\Ek{z} \equiv \frac{1}{2} \rho v_z^2$).  Clearly 
the transverse kinetic energy grows rapidly during the linear development of
the instability but $\Ek{z}$ also grows and, eventually, even becomes larger 
than $\Ek{x}$.  Figure \ref{fig:total-e} contains plots of the total energy 
(kinetic and magnetic) in full-low-hr, hall-med-hr and hall-high-hr.  Somewhat
surprisingly we find that the hall-med-hr simulation loses energy
somewhat {\em faster} than full-low-hr.  This is, on the face of 
it, a little puzzling since the Hall effect does not itself dissipate 
magnetic energy.  If we consider the hall-high-hr simulation we find that the 
total energy is roughly constant - it behaves roughly the same as 
hd-zero-hr.  In fact, the magnetic energy in the hall-high-hr 
simulation behaves qualitatively differently to the hall-med-hr.  Figure 
\ref{fig:eb-hall} contains plots of the perturbed 
magnetic energy as a function of time for hall-high-hr, and full-low-hr.  It 
is clear that something dramatic is happening in the hall-high-hr simulation.  
To gain insight into this, in figure \ref{fig:eb-hall-high-z} we plot the 
total perturbed magnetic energy and the magnetic energy in the $xy$-plane and 
in the $z$-direction for simulation hall-high-hr.  The growth of the magnetic 
field in the $z$ direction is significant, as we expect from a system with the 
Hall effect.  In fact, the growth of the magnetic field in the $xy$-plane is 
faster in the hall-med-hr case.  This isn't too surprising as, in order to 
increase the Hall effect, the coupling between the electrons and the neutrals 
is much weaker than that between the ions and neutrals.  Hence, while the
neutrals and ions generate the usual cat's eye vortex, the electrons
(and the magnetic field) do not.  Figure
\ref{fig:hall-neutral-ion-electron} demonstrates these morphological
differences between the various fluids.

\begin{figure}
\centering
\includegraphics[width=8.4cm]{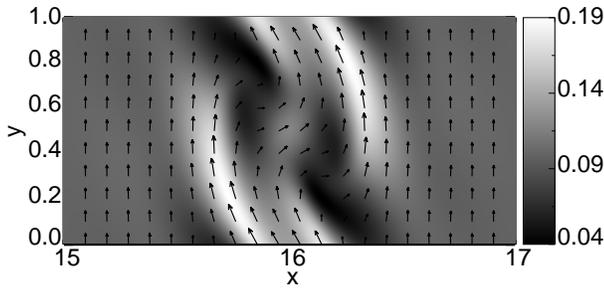}
\caption{Plot of the magnitude and vector field of the magnetic field in the 
$xy$-plane in hall-high-hr at time $8\,t_{\rm s}$. It can be seen that the 
magnetic field does not undergo as much wind-up as is seen in the bulk 
velocity field (upper panel of figure \ref{fig:hall-neutral-ion-electron}).} 
\label{fig:hall-b-field}
\end{figure}

\begin{figure}
\centering
\includegraphics[width=8.4cm]{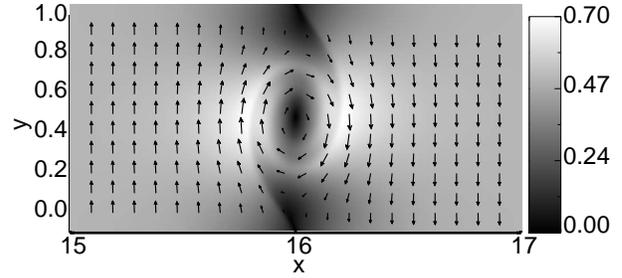}
\vspace{0.1cm}
\includegraphics[width=8.4cm]{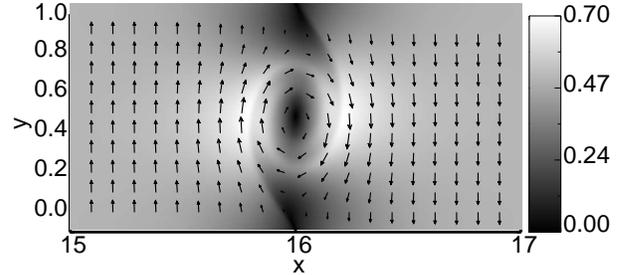}
\vspace{0.1cm}
\includegraphics[width=8.4cm]{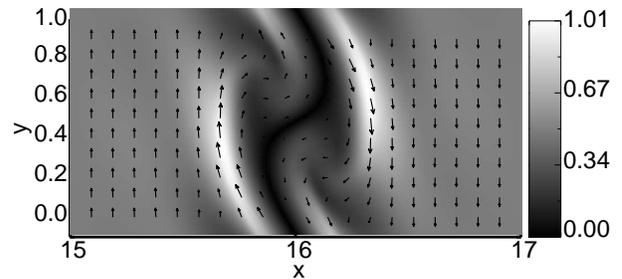}
\caption{Plot of the magnitude and vector field of the bulk velocity (upper 
panel), ion velocity (middle panel) and electron field (lower panel) for
hall-high-hr. The dynamics of the electron fluid can be seen to be very 
different to the other two fluids, and more similar to the magnetic field 
(figure \ref{fig:hall-b-field}).} 
\label{fig:hall-neutral-ion-electron}
\end{figure}

Furthermore, in the case of full-low-hr, the magnetic field in the
$xy$-plane grows to such an extent that it opposes further wind-up of
the fluid in the vortex.  In the hall-high-hr case this does not happen as
magnetic energy is re-distributed to the $z$ component of the field
which does not oppose this wind-up.  Hence the vortex is not destroyed
in the non-linear regime in the hall-high-hr case.  This is the main difference 
between the hall-med-hr and hall-high-hr simulations.  In the non-linear
regime, then, the ion fluid remains spinning in the KH vortex while the
electron fluid remains tied to the magnetic field.  This maintains a
velocity difference between the electrons and ions which, in turn,
causes the magnetic energy to be further re-distributed to the $z$ 
component.  In this way the Hall effect, if strong enough, introduces strong 
dynamo action into the KH instability. This dynamo behaviour,
which is not possible in a 2.5D, ideal MHD system is made possible by the 
Hall term introducing a handedness into the flow \citep[e.g.][]{wardle99disk}.

\begin{figure}
\centering
\includegraphics[width=8cm]{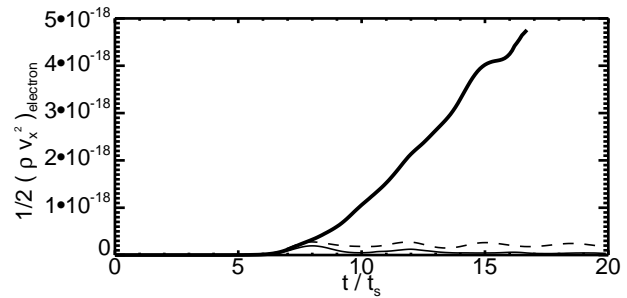}
\caption{The transverse kinetic energy of the electron fluid is plotted 
against time for full-low-hr (thin line) and hall-high-hr (thick line). Also 
plotted in the electron transverse kinetic energy for hd-zero-hr (dashed 
line).} 
\label{fig:ek-electron}
\end{figure}

If we follow the dynamics further we find that the KH instability, in
the presence of high Hall resistivity does not saturate to a
quasi-steady state as it does in, for example, the full-low-hr case.
As the $z$ components of the current and magnetic field continue to grow,
the Hall effect now acts on the non-parallel currents and magnetic fields that
have arisen between the $z$-directions and the $xy$-plane.  This has the 
result of re-orienting some of the magnetic field, and electron fluid flow,
back onto the $xy$-plane. During this process the electron fluid obtains a 
velocity away from the KH vortex, which results in a broader volume of 
plasma being disturbed. This feeds the continuous growth of the magnetic 
energy in the $xy$-plane, and thus causes continuous growth of the
electron transverse kinetic energy, as can be seen in figure
\ref{fig:ek-electron}. In this particular simulation, the size of the simulation grid is set so that it provides a sufficiently large region of ordered flow that can be transformed into disordered flow during the time of the simulation. The instability will naturally saturate when it has exhausted the area of ordered flow available to it.

To summarise, through their strong decoupling from the magnetic field, the 
dynamics of the bulk fluid and ion fluid demonstrate behaviour very
similar to that of hd-zero-hr in which the KH vortex 
remains intact.  The high Hall case can in fact be thought of as two 
separate systems occurring simultaneously; the bulk fluid demonstrating 
hydrodynamic behaviour and the continuously widening volume of perturbed fluid 
perpetually feeding the growth of the magnetic field through the Hall effect. 
These two systems are intrinsically entwined through the requirement of 
charge neutrality, by which the electron fluid causes a widened area of 
perturbed ion fluid, and thus the bulk fluid. Both systems are relatively 
energy efficient and, following the initial growth of the KH instability, the 
overall system experiences little further loss of energy.  The supply of 
energy to feed the magnetic field is limited only to the physical size of the 
computational domain over which the simulation is run. 

This dynamo action occurs only under certain conditions. As
the initial Hall resistivity is increased from moderate to high, the
electron fluid, and thus the magnetic field, becomes increasingly
decoupled from the neutrals. As a result, the neutral fluid tends
toward hydrodynamic behavior. This dynamo action is observed only
when the Hall resistivity is increased to a sufficiently high
value that the KH vortex formed by the neutral fluid is no longer
constrained by the magnetic field. Unlike the pure hydrodynamic
case, the presence of charged fluids undergoing different dynamics
leads to the dynamo action observed. If the Hall resistivity is
not sufficiently large, the magnetic field eventually leads to the
destruction of the KH vortex in the neutral fluid through reconnection as in the ideal MHD case.

Even a moderate amount of Hall resistivity results in a wider volume of fluid 
being disturbed by the instability. This agrees with previous studies of the 
Hall effect on the KH instability \citep[e.g.][]{huba94}.  The re-orientation 
of the magnetic field lines within the KH instability has also been observed 
in studies of MRI in accretion disks \citep[e.g.][]{wardle99disk}.
\cite{kunz08} investigated a simple model of accretion disks without
rotation and demonstrated that the combined actions of the shear instability 
and the Hall effect leads to increased stretching of the magnetic field lines. 
This was shown to result in continued growth of the instability, which 
corresponds well to the dynamo action observed in our simulations with high 
Hall resistivity.  It is important to note, though, that studies by both 
\cite{kunz08} and \citet{wardle99disk} are linear studies and, as such, do
not extend to the nonlinear regime of the instability. A more recent numerical
study by \citet{nykyri04} demonstrates the twisting of magnetic field 
lines, but due to the inclusion of magnetic reconnection, doesn't produce
the magnetic dynamo observed here.

%
% Figure fig:ek-ekz is figure 5.12 in ACJ
% Figure fig:total-e is a variant of figure 5.13 and 5.35 in ACJ
% Figure fig:ek-electron is 5.30 in ACJ
%

%
%  Important point is that the instability doesn't really saturate to a
%  quasi-steady state.
%

%
%  For very high Hall we basically decouple the ions from the B-field.
%  What does this do to the ambipolar resistivity?  This'll basically be
%  a rehash of the medium Hall results so plot the lot together ...
%  Finally, the vortex doesn't break up for high Hall and so we get
%  strong dynamo action.
%

\section{Conclusions}

We have presented the results of a suite of fully multifluid MHD
simulations of the KH instability in weakly ionised fluids such as, for
example, molecular cloud material.  Through varying the collision coefficients 
between the various charged species and the neutrals we were able to 
investigate systems in which ambipolar diffusion dominates the multifluid 
effects and ones in which the Hall effect dominates.  We validated our
KH simulations through comparison of an %(effectively) 
ideal MHD simulation with previously published results and performed resolution studies 
for each of these cases to ensure that our conclusions are not unduly
effected by our numerical resolution.

Our findings can be summarised as follows:
\begin{itemize}
\item The multifluid effects do not significantly influence the linear
growth rate of the instability.
\item Ambipolar diffusion dramatically reduces the energy associated
with the perturbed magnetic field.  This happens through diffusion %reconnection
for moderate ambipolar resistivity, 
%diffusion, 
but through both diffusion 
%reconnection 
and decoupling of the magnetic field from the bulk flow for higher resistivity.
%diffusion.
\item The Hall effect, as expected, rotates the magnetic field out of the 
initial $xy$-plane.
\item In contrast to both ambipolar dominated and ideal MHD flows, the
Hall effect causes the system to fail to settle to a quasi-steady state after 
saturation of the instability.  
\item For moderate Hall resistivity the perturbed magnetic field contains 
higher energy than in the ideal MHD case, presumably due to a field topology 
which impedes (numerical) reconnection.
\item For high Hall resistivity strong dynamo action is seen as energy
associated with the magnetic field grows without any apparent signs of 
saturation.
\end{itemize}

\section*{Acknowledgements}
The research of A.C.J. has been part supported by the CosmoGrid project funded 
under the Programme for Research in Third Level Institutions (PRTLI) 
administered by the Irish Higher Education Authority under the National 
Development Plan and with partial support from the European Regional 
Development Fund.

The authors wish to acknowledge the SFI/HEA Irish Centre for High-End
Computing (ICHEC) for the provision of computational facilities and
support.

\label{lastpage}

\end{document}